\documentclass[aps,pra,twocolumn,superscriptaddress,noshowpacs]{revtex4-1}
\usepackage{latexsym}
\usepackage{amsmath}
\usepackage{amssymb}
\usepackage{graphicx}
\usepackage{caption}
\usepackage{subfigure}
\usepackage{float}
\usepackage{mathrsfs}

\newcommand{\so}{\text{s}}
\newcommand{\gt}{\text{g}}
\newcommand{\dr}{\text{d}}

\begin{document}
\title{Atom transistor from the point of view of quantum nonequilibrium dynamics}

\author{Zhedong Zhang}
\affiliation{Department of Physics and Astronomy, State University of New York at Stony Brook, Stony Brook, NY 11794  USA}
\author{Vanja Dunjko}
\affiliation{Department of Physics, University of Massachusetts Boston, Boston, MA 02125, USA}
\author{Maxim Olshanii}
\affiliation{Department of Physics, University of Massachusetts Boston, Boston, MA 02125, USA}
\date{\today}

\begin{abstract}
We analyze the atom field-effect transistor scheme
[J. A. Stickney, D. Z. Anderson and A. A. Zozulya, Phys. Rev. A 75, 013608 (2007)]
using the standard tools of nonequlilibrium dynamics. In particular, we
study the deviations from the Eigenstate Thermalization Hypothesis, quantum fluctuations, and the density of states,
both ab initio and using their mean-field analogues. Having fully established the quantum vs.~mean-field correspondence
for this system, we attempt, using a mean-field model,  to interpret the off-on threshold in our transistor
as the onset of ergodicity---a point where the system becomes able to visit the thermal values of the
former integrals of motion in principle, albeit not being fully thermalized yet.
\end{abstract}

\maketitle

\section{Introduction}
Similarly to the conventional electronic transistors, in atomtronics, a transistor is a nonlinear
device where a small atom current or atom number controls a large current. The existing proposals for an atom transistor can
be divided into open and closed architectures. The former \cite{seaman2007_023615,pepino2009_140405,pepino2010_013640} is
conceptually close to the conventional electronic devices (the work \cite{pepino2009_140405} also contains a design for a diode). The latter, in turn,
can be separated into the schemes based on an adiabatic population transfer protocol
\cite{vaishnav2008_265302,benseny2010_013604,schlagheck2010_065020} (along with the similar atom diode proposals
\cite{ruschhaupt2004_061604,ruschhaupt2006_3833}),
and the schemes where the (large) base current is induced by the difference
in chemical potentials between the ``electrodes''.
The base current can be controlled by either the internal state of a localized impurity
\cite{micheli2004_140408,daley2005_043618,micheli2005_0506498} or
by the number of (strongly interacting) atoms in the middle site \cite{stickney2007_013608,caliga2013_12083109}. This latter scheme is nothing else but
an atomic version of the \textit{field-effect transistor} (FET).
Here, the large atomic current from the ``source'' to the ``drain'' is controlled by a a small number of atoms in the ``gate.'' The interatomic repulsion in the gate is so strong that the chemical potential is large even when the number of atoms in the gate is small.

Here we show that the off-on threshold in our transistor
corresponds to the onset of ergodicity---the point where the system becomes able to visit the thermal values of the
former integrals of motion in principle, albeit not being fully thermalized yet.

\section{System of interest}
The paradigmatic atom transistor is realized in a tight binding model for three coupled bosonic wells on a line, the ``source'', the ``gate'', and the ``drain''
\cite{stickney2007_013608}. The Hamiltonian of the system reads
\begin{align}
\begin{split}
\hat{H} =
& \sum_{\eta=\so,\gt,\dr}\left(\varepsilon_{\eta}b_{\eta}^{\dagger}b_{\eta}+\frac{U_{\eta}}{2}b_{\eta}^{\dagger}b_{\eta}^{\dagger}b_{\eta}b_{\eta}\right)
\\
&+ J_{\so\gt}\left(b_\so^{\dagger}b_\gt+b_\gt^{\dagger}b_\so\right) + J_{\gt\dr}\left(b_\gt^{\dagger}b_\dr+b_\dr^{\dagger}b_\gt\right)\,,
\end{split}
\label{H}
\end{align}
where the indices ``\so'', ``\gt'', and ``\dr'' stand for source, gate and drain, respectively, $\varepsilon_{\eta}$ and $U_{\eta}$ are
the site-dependent one-body energy and the strength of the on-site two-body interactions for a site
$\eta = \so,\, \gt,\,\dr$, and $J_{\eta\eta'}$ is the hopping constant between sites $\eta$ and $\eta'$ .

In general, the purpose of the atom FET transistor is to control the current of atoms from
the source to the drain using small variations in the chemical potential of the gate, $\mu_{\gt} = \varepsilon_{\gt} - U_{\gt} n_{\gt}$,
controlled in turn by its population, $n_{\gt}$. According to the original proposal \cite{stickney2007_013608}, the transition of atoms between the
source and the gate plays the role of a ``bottleneck,'' activated only when the source and gate chemical potentials
become close; in contrast, the gate-drain link is made insensitive to the gate-drain chemical potential difference, transmitting every atom that
happens to appear at the gate.
This can be achieved by providing a comparatively large on-site interaction strength for the gate, subsequently detuned from the energy
of the source. One then chooses the hopping constants and the on-site energies in such a way that the source-gate transition exhibits a
narrow resonance, and the gate-drain transition a broad one. We will be using the following representative set of parameters: $J_{\so\gt}=-0.1$,
$U_\gt N=100$, $\varepsilon_\gt=-1.3$, $J_{\gt\dr}=-1.0$, $\varepsilon_\dr=0.5$ and $N=30$, with the rest of the parameters set to zero:
$U_\so N = \varepsilon_\so = U_\dr N = 0$.

\begin{figure}[H]
\textbf{(a)}
\includegraphics[width=0.4\textwidth]{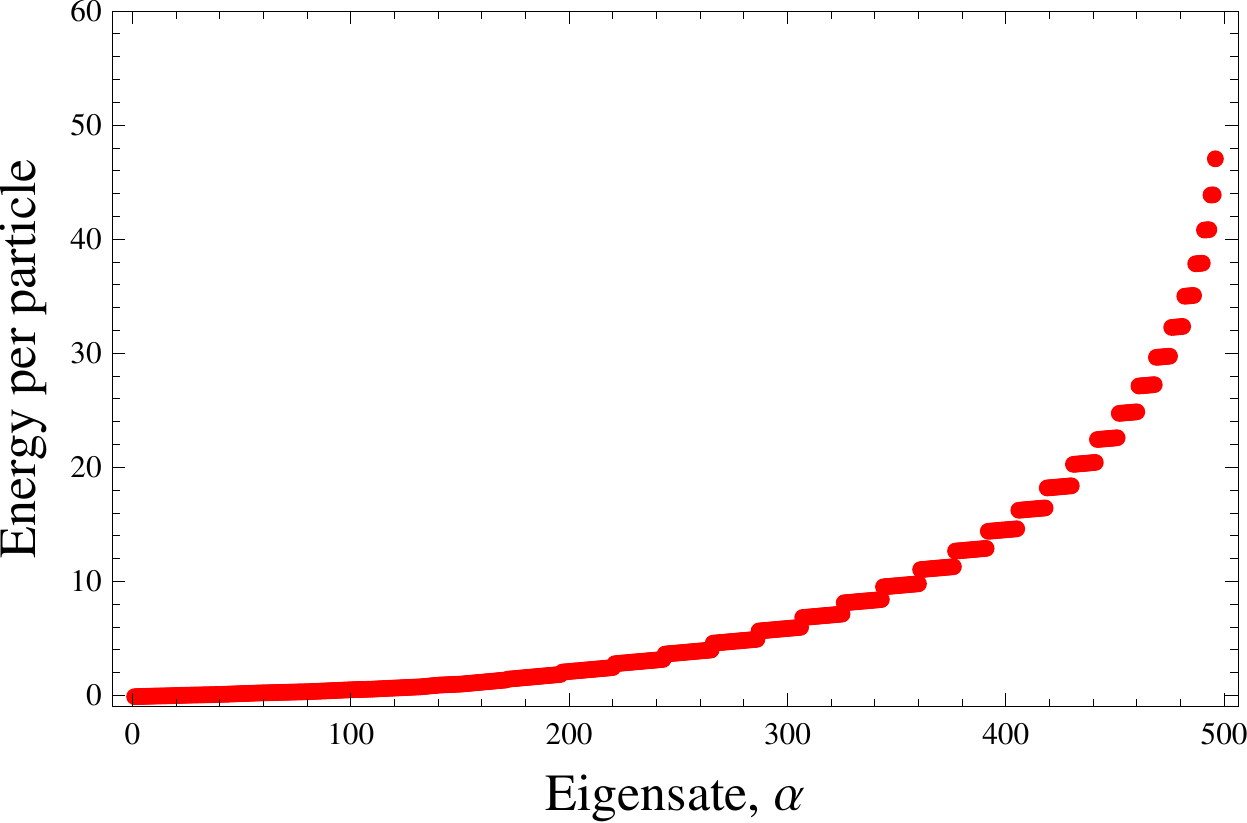}
\\
\textbf{(b)}
\includegraphics[width=0.4\textwidth]{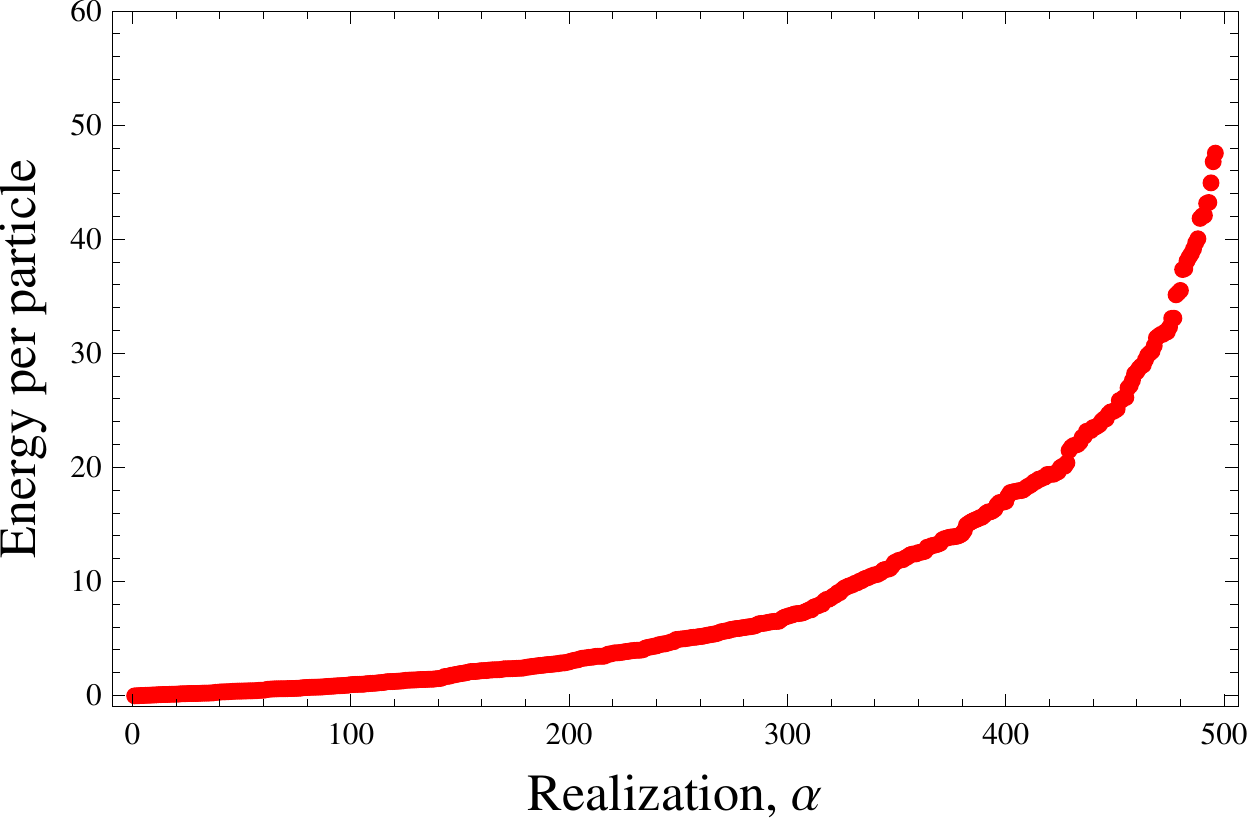}
\caption{(color online) Energy spectrum of the atom transistor. Parameters are $J_{\so\gt} = -0.1$, $U_{\gt} N = 100.$, ${\cal E}_{\gt} = -1.3$, $J_{\gt\dr} = -1.$, ${\cal E}_{\dr} = 0.5$, while $U_\so N = \varepsilon_\so = U_\dr N = 0$.
{\bf (a)} The ab initio energy spectrum ($496$ eigenstates) for an atom transistor with $N=30$ atoms.
{\bf (b)} The mean-field emulation of the quantum energy spectrum. 496 initial conditions were uniformly distributed through the whole available phase space. The vertical axis gives the temporal average of total energy, for each realization.
The realizations were permuted in such a way that the energies increase along the list. The vertical axis
reflects the position of the realizations in this list.
}
\label{energy__both}
\end{figure}

Figure~\ref{energy__both}a shows the energy spectrum $E_{\alpha}$ of the transistor for the chosen set of parameters. Here and below,
$\alpha$ is the index of the eigenstate $|\alpha \rangle$ of the Hamiltonian in Eq.~(\ref{H}): $\hat{H} |\alpha \rangle = E_{\alpha} |\alpha \rangle$. The spectrum,
bounded from both below and above, contains 496 eigenstates. Note that the density of states decreases with energy.
On the upper end of the spectrum, the dominant contribution to the energy is provided by the
interactions between the atoms in the gate, $n_\gt$; this energy increases quadratically with $n_\gt$,
leading to the decreasing (with energy)
energy spacings.

Most of the spectrum turns out to be within the limits of applicability of the semiclassical approximation.
To verify this assertion, we check that Weyl's law holds in our system. According to Weyl's law, if one computes how the number of quantum states below an energy $E$
\noindent depends on the energy $E$, and then takes the smooth envelope of this dependence, the result is proportional to the classical phase space volume occupied by these states:
%
\begin{multline}
(\mbox{\# of states with } E_{\alpha} \le E) 
\\
\approx
    									       \frac{1}{(2\pi\hbar)^3}
                                                                                   \int_{0}^{2\pi} \! d \phi_\so
                                                                                   \int_{0}^{\infty}\!  d I_\so
                                                                                   \int_{0}^{2\pi} \!  d \phi_\gt
                                                                                   \int_{0}^{\infty}\!  d I_\gt
                                                                                   \int_{0}^{2\pi}\! d \phi_\dr
                                                                                   \\ \times
                                                                                   \int_{0}^{\infty}\! d I_\dr \,
                                                                                   \Theta(E-H(\phi_\so,\, I_\so,\, \phi_\gt,\, I_\gt,\, \phi_\dr,\, I_\dr))
\,\,,
\label{Weyl's_law}
\end{multline}
%
where
\begin{multline*}
%
H(\phi_\so,\, I_\so,\, \phi_\gt,\, I_\gt,\, \phi_\dr,\, I_\dr) =
 \sum_{\eta=\so,\gt,\dr}\left(\varepsilon_{\eta}I_{\eta}+\frac{U_{\eta}}{2}I_{\eta}^2\right)
\\
+ 2 J_{\so\gt} \sqrt{I_\so I_\gt} \cos[\phi_\so-\phi_\gt] + 2 J_{\gt\dr} \sqrt{I_\gt I_\dr} \cos[\phi_\gt-\phi_\dr]
\label{H_classical}
\end{multline*}
is the corresponding mean-field Hamiltonian, expressed through the action-angle variables of the zero-hopping analogue of the Hamiltonian in Eq.~(\ref{H}),
%
%
$
I_{\eta} = \langle b_{\eta}^{\dagger}b_{\eta} \rangle
$,
$
\mbox{$\phi_{\eta}-\phi_{\eta'}$} = \arg[\langle b_{\eta}^{\dagger} b_{\eta'} \rangle]
$,
%
%
and $\Theta(x)$ is the Heaviside step-function. Figure~\ref{energy__both}b shows the inverse of the dependence (\ref{Weyl's_law}), computed using a Monte-Carlo method:
496 phase-space points were distributed uniformly within an intersection of the whole available volume of the phase space and a narrow shell corresponding
a constant window of the values of the norm (where
$
\mbox{norm} \equiv
                                                                                   \int_{0}^{2\pi} \! d \phi_\so
                                                                                   \int_{0}^{\infty}\!  d I_\so
                                                                                   \int_{0}^{2\pi} \!  d \phi_\gt
                                                                                   \int_{0}^{\infty}\!  d I_\gt
                                                                                   \int_{0}^{2\pi}\! d \phi_\dr
                                                                                   \int_{0}^{\infty}\! d I_\dr \,
                                                                                   (I_\so+I_\gt+I_\dr)
$), around $\mbox{norm} = N$. Notice that for this calculation, we have chosen as many Monte-Carlo realization as there are quantum eigenstates of the system.
A priori, one number is not related to the other, and more realizations would produce a smoother semi-clalssical curve. However, it can be argued that
Fig.~\ref{energy__both}b can serve as a semi-calssical \textit{emulation} of the quantum spectrum Fig.~\ref{energy__both}a, where the fluctuations of the
eigenenergies around a smooth envelope have at least the same order of magnitude as their
semiclassical counterparts. Indeed, on the one hand, the Monte-Carlo method
produces
energy values through a Poisson process; on the other hand, it is known that in quantum integrable systems, eigenenergies are distributed as if they are the outcome of a Poisson process \cite{bohigas1991,guhr1998}.
In contrast, quantum-ergodic systems with time-reversal invariance exhibit the type-1 Wigner-Dyson statistics \cite{bohigas1991,guhr1998}. In that case, the standard deviation for the spacing between two neighboring levels is $1/\sqrt{(4/\pi) - 1} \approx 2$ times lower than in the case of Poisson statistics.

\section{Atom transistor from the point of view of quantum nonequilibrium dynamics: the off-on crossover as the point at which the system becomes able to reach thermal equilibrium values}
\begin{figure}[H]
\textbf{(a)}
\includegraphics[width=0.4\textwidth]{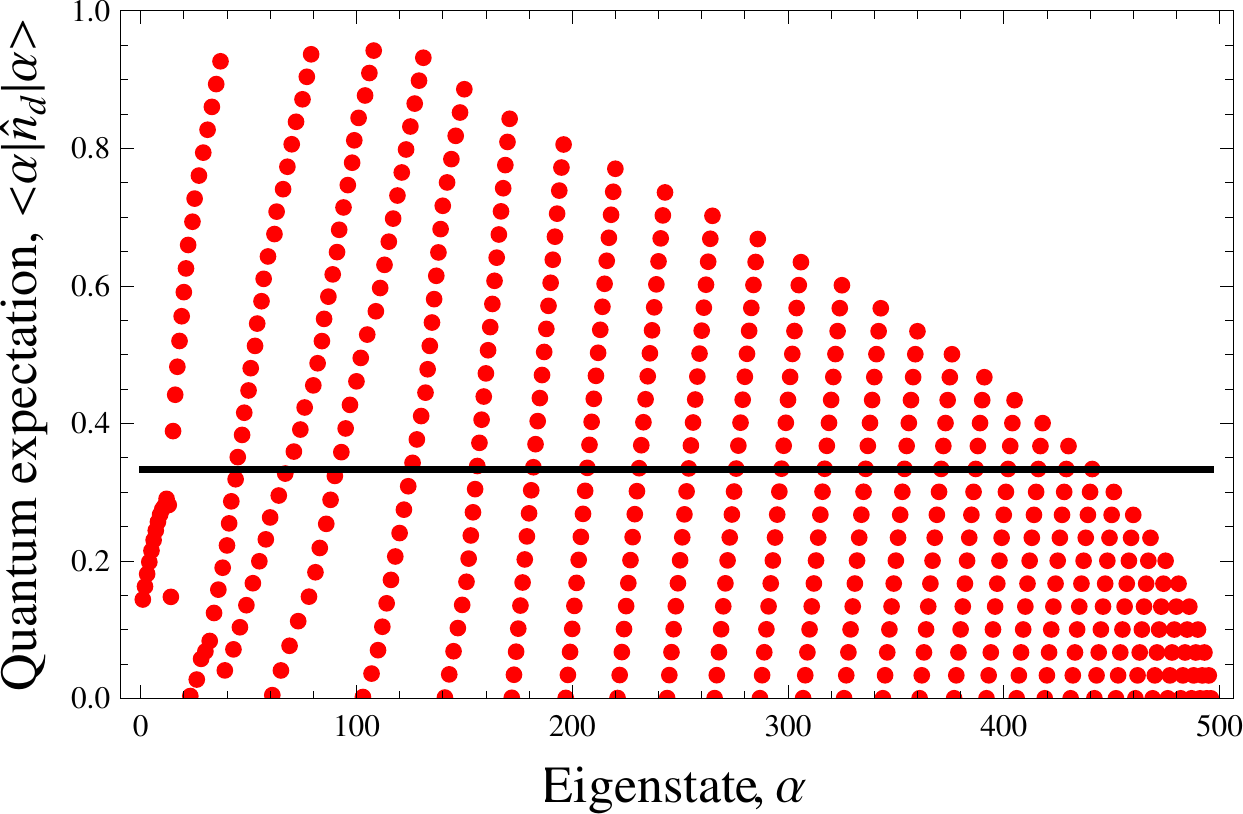}
\\
\textbf{(b)}
\includegraphics[width=0.4\textwidth]{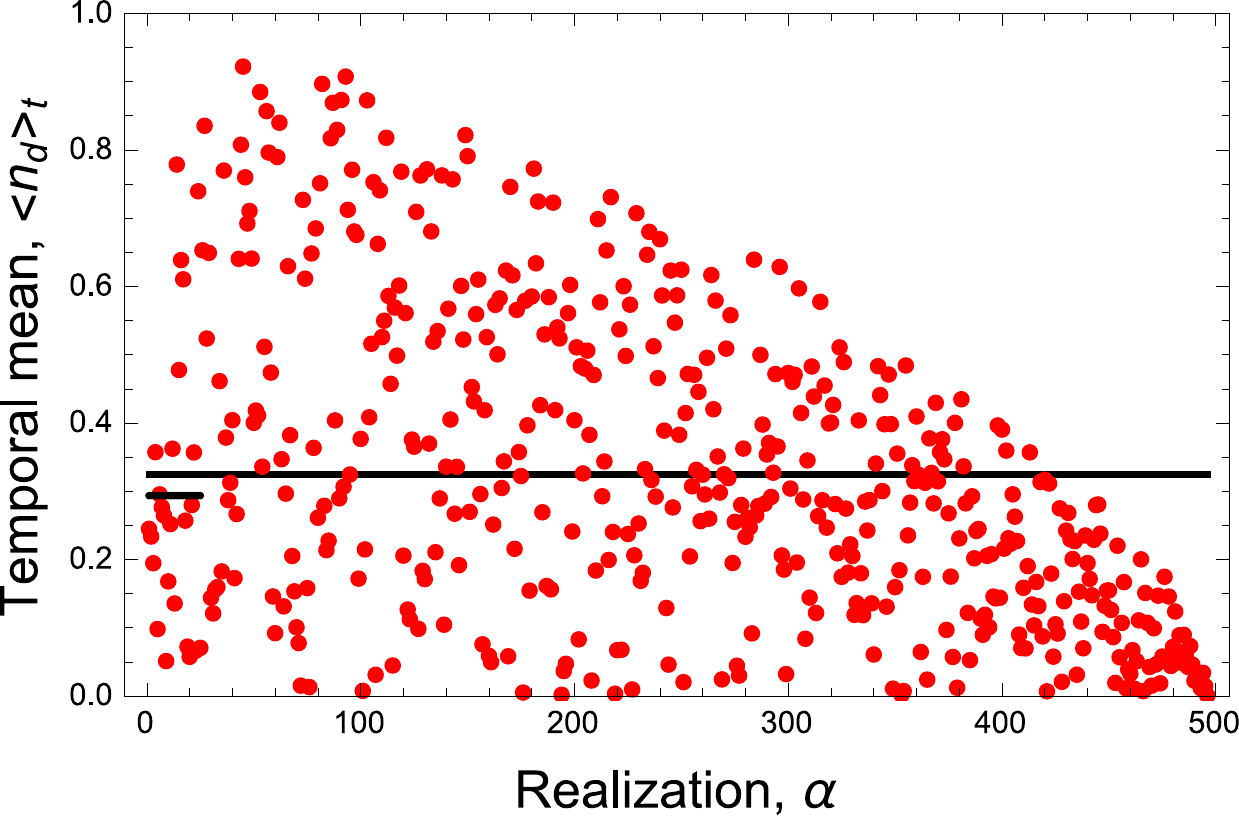}
\caption{(color online)
Expectation values of the drain occupation. {\bf (a)} Quantum expectation values in the eigenstates.
The solid horizontal line reflects the average over all the eigenstates.
The system parameters are the same as in Fig.~\ref{energy__both}(a).
{\bf (b)} Time average of the drain occupation of the atom transistor, in the mean-field approximation, for all the realization used. The long solid horizontal line reflects the average over all the realizations. The short solid horizontal line
is the average over 25 realizations with the lowest energies;
these occupy approximately 1/20 of the total available phase space volume. This portion of
the phase space constitutes the domain of operation of the atom transistor (see Figs.~\ref{detune_chem} and
\ref{gain_mean_nd}).
The system parameters are the same as Fig.~\ref{energy__both}(b)}.
\label{mean_nd_both}
\end{figure}

Figure~\ref{mean_nd_both}a explores the central object in the eigenstate thermalization theory \cite{deutsch1991,srednicki1994,rigol2008,dunjko2013_AMO_review}---the expectation values of the relevant observables; in our case, this is the relative occupation of
the drain:
%
%
$
\langle \alpha | \hat{n}_\dr | \alpha \rangle
$.
%
%
Notice that this expectation value, when plotted as a function of $\alpha$, does not collapse to a single line; this means our system has not reached eigenstate thermalization. Let
%
%
$
n_{\eta} \equiv N_{\eta}/N
$, where
$
N_{\eta} = I_{\eta} = \langle b_{\eta}^{\dagger}b_{\eta} \rangle
$.
%
%
Figure~\ref{mean_nd_both}b shows the classical infinite time averages of the drain occupation,
\begin{align*}
\langle n_\dr \rangle_{t} \equiv \lim_{t_{\text{max}} \to \infty} \frac{1}{t_{\text{max}}} \int_{0}^{t_{\text{max}}} \! dt \, n_\dr(t)
\,\,,
\end{align*}
where the pseudo-eigenstates of Fig.~\ref{energy__both}b
are used as the initial conditions. The overall behavior of the quantum and classical expectation values stand in good correspondence
 (see comment 24 in Ref.~\cite{olshanii2015_060401}).

\begin{figure}[H]
\textbf{(a)}
\includegraphics[width=0.4\textwidth]{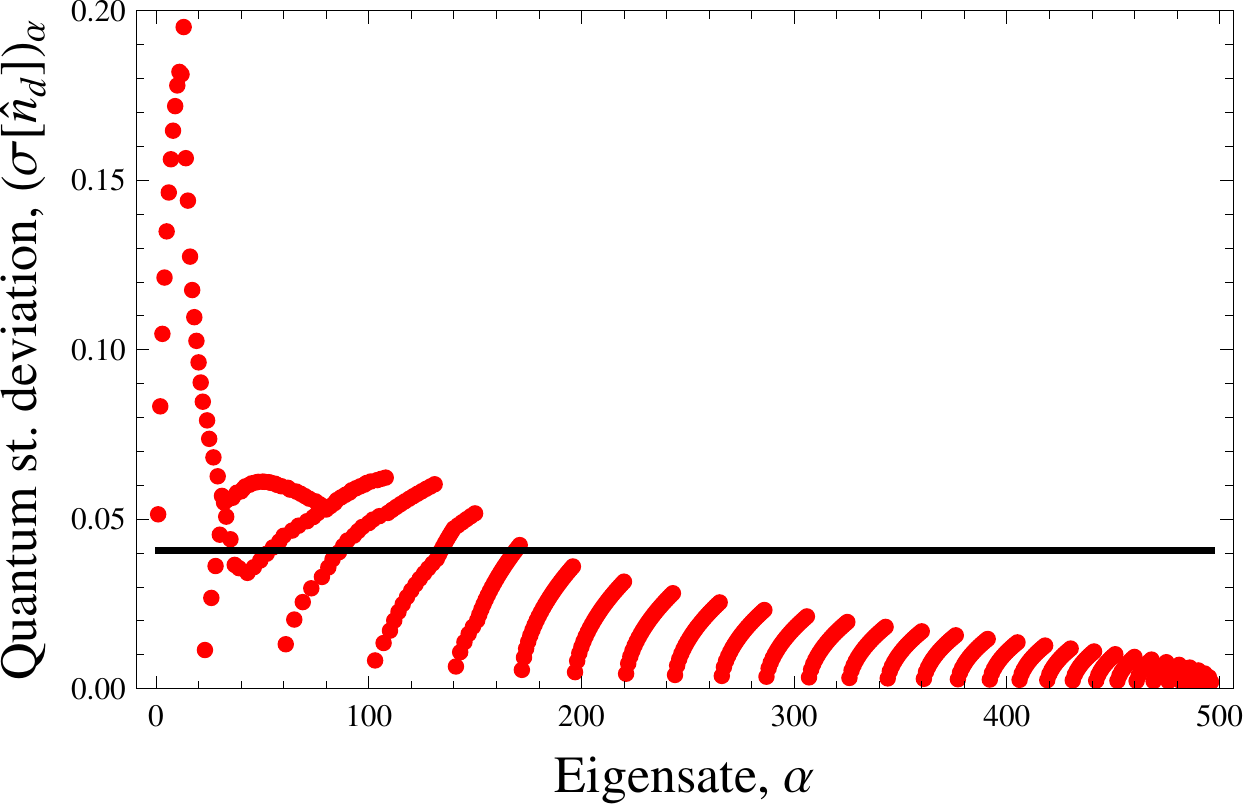}
\\
\textbf{(b)}
\includegraphics[width=0.4\textwidth]{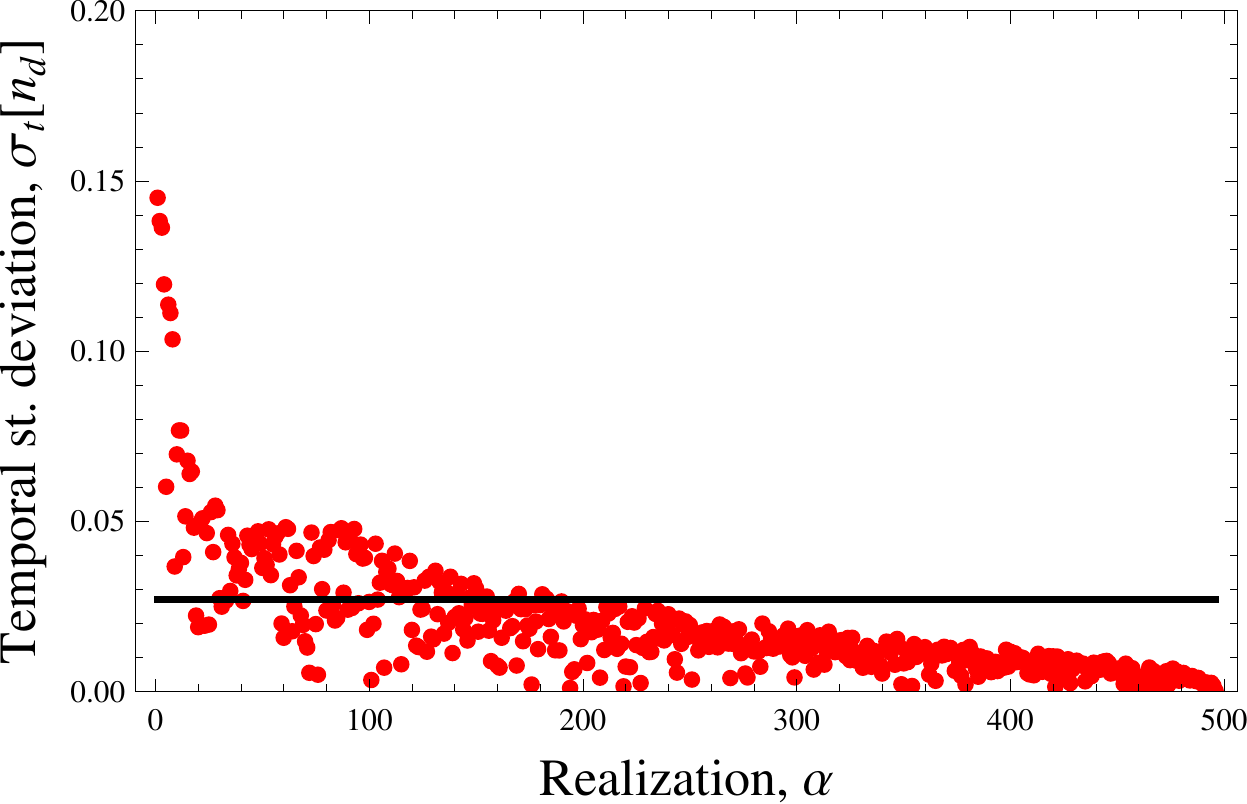}
\caption{(color online) Fluctuations of the drain occupation.
{\bf (a)}
Quantum fluctuation in the eigenstates. The solid horizontal line reflects the average over all the eigenstates. The parameters are the same as in Fig.~\ref{mean_nd_both}(a).
{\bf (b)} Temporal variance for the initial states generated, in the mean-field approximation.
The solid horizontal line reflects the average over all realizations. The parameters are the same as in Fig.~\ref{mean_nd_both}(b)}
\label{StDev_nd_both}
\end{figure}

Quantum fluctuations of $n_\dr$ in the eigenstates $| \alpha \rangle$,
\begin{align*}
\sigma[\hat{n}_\dr ]_{\alpha} \equiv \sqrt{\langle \alpha | (\hat{n}_\dr)^2 | \alpha \rangle  - (\langle \alpha | \hat{n}_\dr | \alpha \rangle)^2}
\,,
\end{align*}
are shown in Fig.~\ref{StDev_nd_both}a.
In accordance with the conjecture
\noindent expressed in Ref.~\cite{srednicki1996_L75}, the classical temporal variance,
\begin{align*}
\sigma_{t}[n_\dr] \equiv \sqrt{\left( \lim_{t_{\text{max}} \to \infty} \frac{1}{t_{\text{max}}} \int_{0}^{t_{\text{max}}} \! dt \, (n_\dr(t))^2  \right) -  (\langle n_\dr \rangle_{t} )^2}
\,,
\end{align*}
shown in Fig.~\ref{StDev_nd_both}b, constitutes
a faithful classical analogue of the quantum fluctuations.

\begin{figure}[H]
\centering\includegraphics[width=0.4\textwidth]{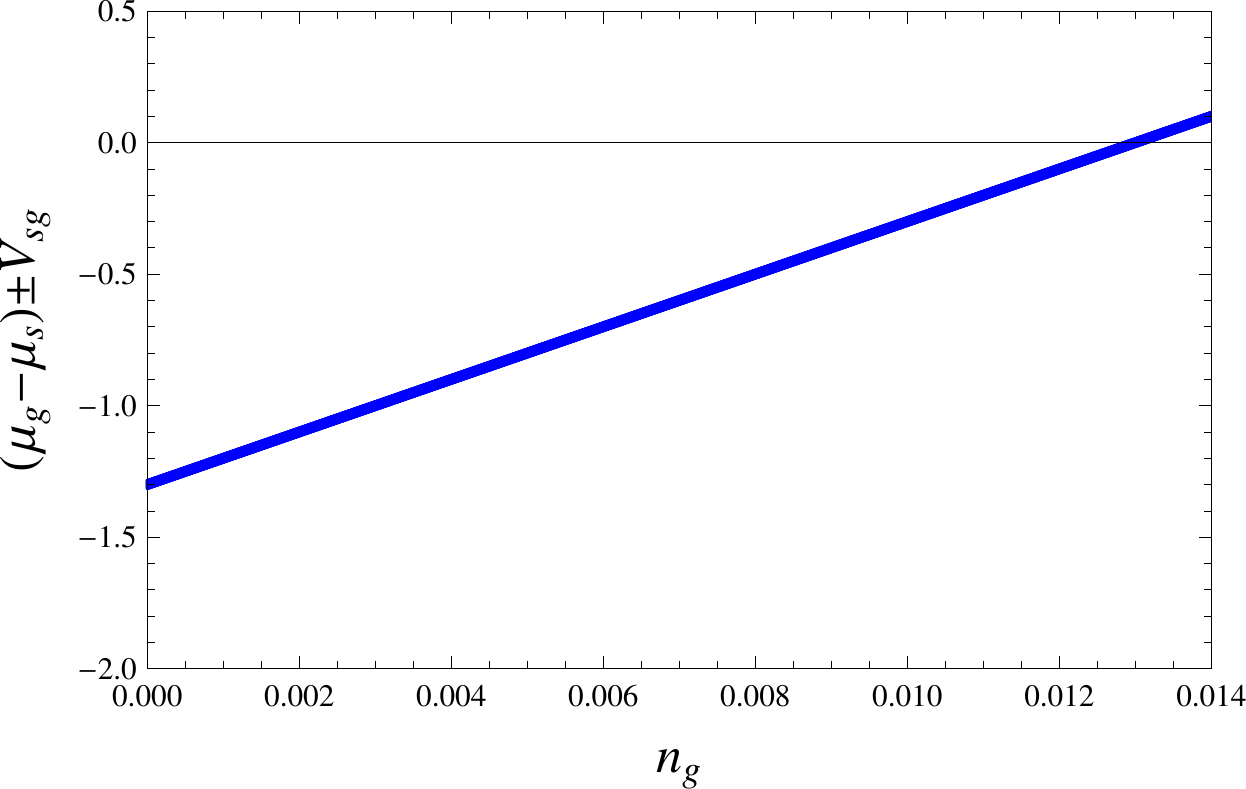}
\caption{(color online)
Detuning between the gate and the source chemical potentials, as a function of the
relative occupation of the gate, $N_{\gt}/N$. The vertical half-width of the line
equals the coupling constant $V_{\so\gt}$ between source and the gate.
The rest of the parameters is the same
as st Fig.~\ref{mean_nd_both}(b).}
\label{detune_chem}
\end{figure}

Figure~\ref{detune_chem} shows the detuning of the source-gate transition as a function of the relative gate occupation.
The resonance occurs at $n_\gt=0.013$. Recall that in our case, $\mu_{\so} = \varepsilon_\so$ and $\mu_{\gt} = \varepsilon_{\gt} - U_{\gt} n_{\gt}$.

In all numerical experiments below, the initial value of the drain occupation $n_\dr$ is zero or very close to zero.

\begin{figure}[H]
\textbf{(a)}
\includegraphics[width=0.4\textwidth]{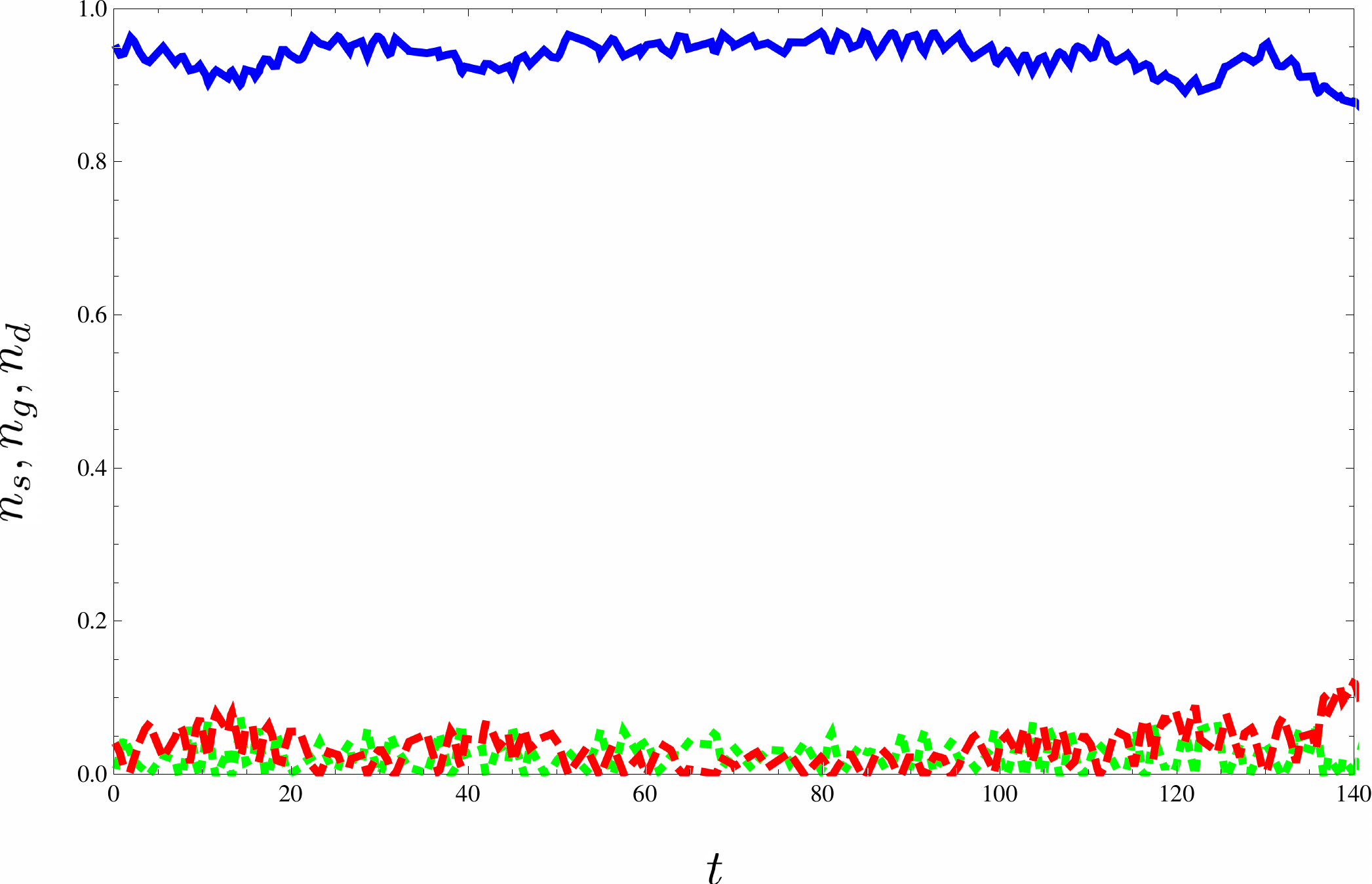}
\\
\textbf{(b)}
\includegraphics[width=0.4\textwidth]{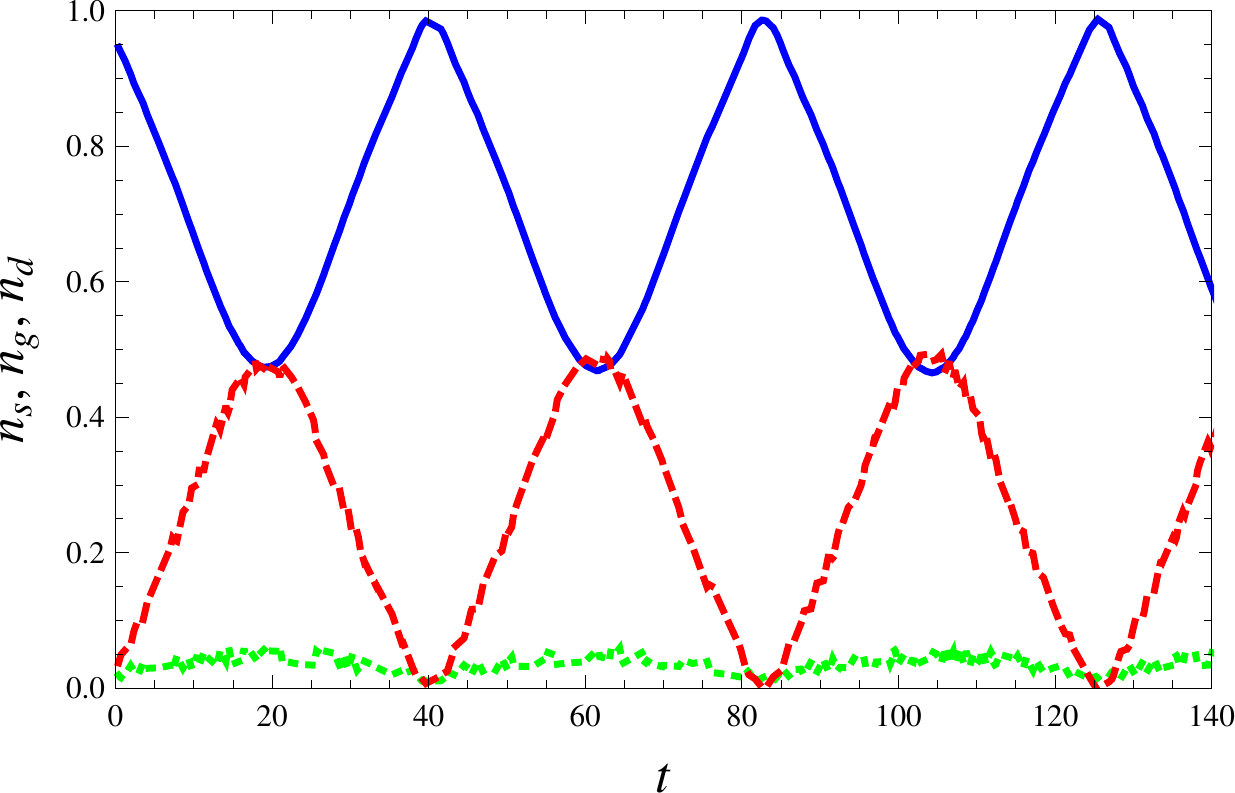}
\caption{(color online) Dynamics of the source (solid, blue), gate (short-dashed, green) and drain (dashed, red) occupations
in the ``on'' and ``off'' modes.
 {\bf (a)} ``Off'' mode (low values of $n_\gt\equiv N_\gt/N$ in Figs.~\ref{detune_chem} and \ref{gain_mean_nd}).
 {\bf (b)} ``On'' mode (value $n_\gt\equiv N_\gt/N$ close to a resonant value of 0.013, according to Figs.~\ref{detune_chem} and \ref{gain_mean_nd}).
 }
\label{transistor_on_off}
\end{figure}

At zero $n_\gt$ initially, only a minimal conductance is expected, and this is what we observe (Fig.~\ref{transistor_on_off}a).
To the contrary, at the resonant point, $n_\gt(t\!=\!0)=0.013$, the transistor is supposed
to provide the maximal source-drain conductance, and this is indeed what we find (Fig.~\ref{transistor_on_off}b).

The conventional figure of merit is the current to the drain, characterized by the initial slope in the dependence of $n_{\gt}$ vs.\ time  (in our case, the slope of the red line between $t=0$ and $t \approx 20$, in Fig.~\ref{transistor_on_off}b). Below, we will use an integral figure of merit---the infinite time average of 

\noindent the drain occupation, $\langle n_\dr \rangle_{t}$---to facilitate the analysis of the system from the point of view of ergodicity or its absence.

\begin{figure}[H]
\centering\includegraphics[width=0.4\textwidth]{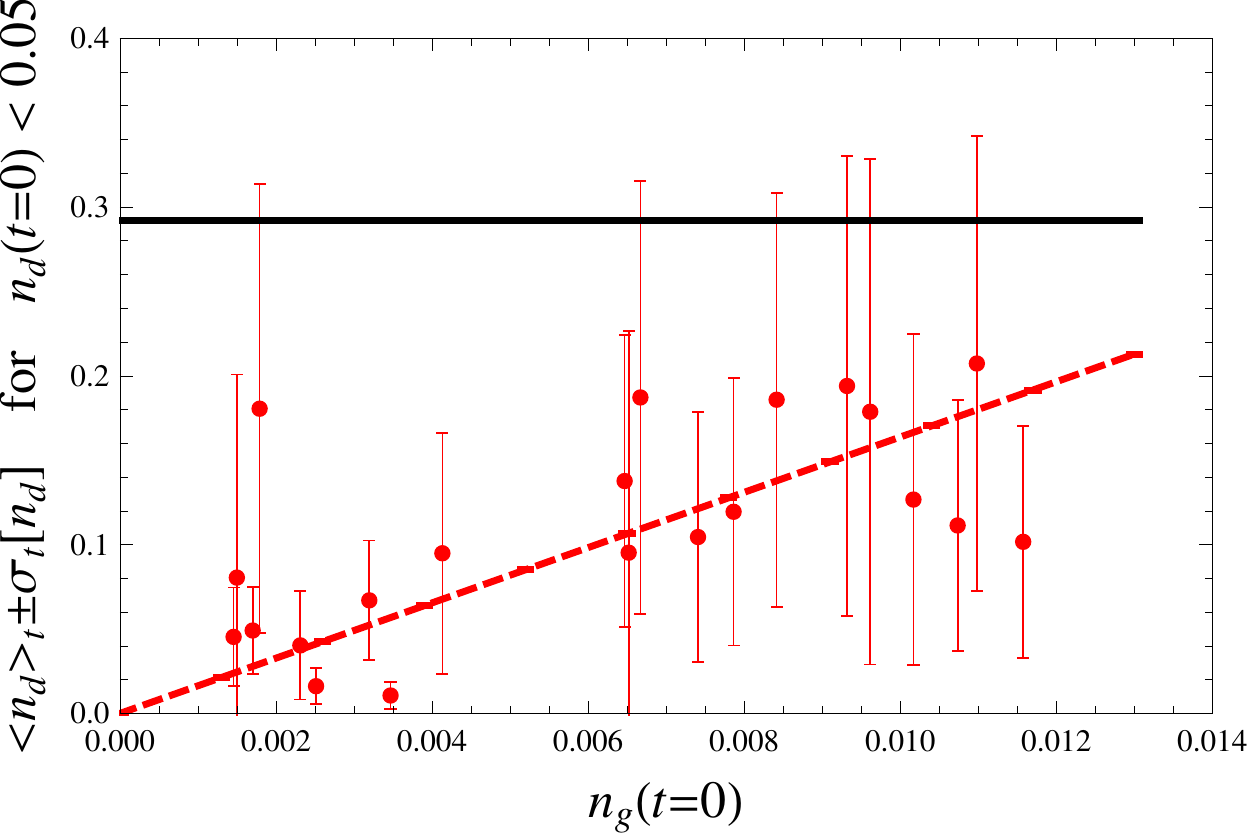}
\caption{(color online) Time average of the drain occupation of the atom transistor (dashed line), along with its temporal
standard deviation (error bars), in the mean-field approximation, as a function of the initial occupation of the gate. 1015 initial conditions were uniformly distributed inside a phase space volume corresponding to the lowest (in energy) 1/20 of the total available phase space volume. The corresponding window of energies is between $E_{\text{min}}/N\approx 0.08$ and $E_{\text{max}}/N\approx 0.13$, out of the full available energy range of $E_{\text{min},\text{full}}/N=E_{\text{min}}/N\approx 0.08$ and $E_{\text{max},\text{full}}=47$. Subsequently, only the initial conditions corresponding to low initial conditions ($N_\dr(t=0)\le 0.05N$) of drain occupation were selected. A linear fit (dashed line) gives an estimate of the gain, $\beta\equiv\langle N_\dr\rangle_t/N_\gt(t=0)$, as $\beta\approx 16$. Solid horizontal line reflects the average over \textit{all} the realizations in the window, with no selection of the initial values of the drain occupation $N_\dr$.
Note that only when the initial gate occupation approaches the resonant value
of $n_{\gt} = .013$ (see Fig. \ref{detune_chem}), the mobility in the phase space becomes sufficient
for the drain occupation to reach its thermal value (solid line), at least at some instances of time.
The system parameters are the same as in Fig.\ref{mean_nd_both}(b)}
\label{gain_mean_nd}
\end{figure}

In Fig.~\ref{gain_mean_nd}, we trace the $\langle n_\dr \rangle_{t}$ vs.\ $n_\gt(t\!=\!0)$ dependence, along with the temporal variation of
$n_{\dr}$. The slope of this dependence, $\beta \approx 16$, we identify as gain. Unexpectedly, we find that even at the resonant point, the
average drain occupation does not reach its ensemble average (also represented by the short horizontal line in Fig.~\ref{mean_nd_both}). Instead, as the system approaches the resonant point $n_\gt(t\!=\!0)=0.013$, the ensemble average starts falling \textit{within the range} of the temporal fluctuations; not enough to be predominantly within the range, but enough for this effect to be detectable.

\section{Conclusion}
In this work, we studied the atom FET transistor scheme suggested in Ref.~\cite{stickney2007_013608,caliga2013_12083109}
using the tools of quantum non-equilibrium dynamics \cite{olshanii2015_060401}. We first justified the applicability of the semiclassical
approximation applied to the standard measures of the non-equilibrium. We then focused, using a semiclalssical model (equivalent to a mean-field one), on the initial conditions with zero drain occupation,
and used the gate occupation as the knob that controls further dynamics. Instead of the 

\noindent  traditional source-to-drain current, as the transistor output we choose the infinite time average of the drain occupation: this change brings us closer to the conventional measures of non-equilibrium. We then
studied how the drain occupation depends on the initial gate occupation. At zero gate occupation (transistor ``off''),
no atoms are transmitted to the source. To the
contrary, at a point where the gate's chemical potential levels with that of the source (transistor ``on''),
the drain becomes populated. One may expect that
at this point, the transistor is fully thermalized, i.e. that the infinite time averages of the
observables approach their (microcanonical) ensemble averages.
However, it turned out that even at the resonant point, the drain occupation is still far below its thermal value.
Instead, we found that the ``on'' regime is characterized by the accessibly of the thermal values of the drain in ``principle''. Namely, around the
``on'' value of the initial gate occupation, the temporal fluctuations of the drain occupation become capable of ``touching'' the thermal values
for some periods of time, without spending much time there. The overall conclusion is that the ergodicity is not a necessary condition for
the operation of atom transistor (while the conventional semiconductor devices do operate close to a thermal equilibrium):
this conclusion may allow one to broaden the search for an optimal configuration of an atom transistor.

\acknowledgments{We are grateful to Dana Anderson and Alex Zozulya for numerous in-depth discussions on the subject. This work was supported by grants from National Science Foundation (\textit{PHY-1402249}) and the Office of Naval Research (\textit{N00014-12-1-0400}).}


%

\end{document}